# Radiative View Factor Correlations in Particulate Media from Ray Tracing Simulations and Data-Driven Modeling

Zijie Chen[1] and Rohini Bala Chandran[1*]

[1] *2350 Hayward St., G. G. Brown Building, Department of Mechanical Engineering, University of Michigan, Ann Arbor, MI 48109*

*Corresponding author. Email: rbchan@umich.edu, Phone: 734-647-9049

## Abstract

Thermal radiation in particulate media has been extensively modeled by solving the radiative transport equation with effective radiative properties or with statistical ray tracing techniques. While effective for static particles, this approach is not compatible with the more common discrete formulation of heat fluxes in particle flow systems. This study focuses on such a discrete approach to compute radiative fluxes by developing view factors correlations for particle-particle and particle-wall view factors. Training data is generated from physics-based Monte Carlo ray tracing simulations on a monodisperse, packed bed with solid volume fractions ranging from 0.016 to 0.45. This data was used to develop reduced-order correlations to determine particle-particle and particle-wall radiative view factors as a function of particle-particle and particle-wall separation distance, viewing angle, and the number of shading particles. Uniquely, we determine best-fit functions that are physically interpretable to account for shading effects by particles. A sigmoid function with a non-linear dependence on viewing angle governs the extent of shading cast by an intermediate particle. To scale the net contribution of shading by intermediate particles between a particle and a planar wall, a particle-wall correction factor that is dependent quadratically on the particle-wall normal separation distance serves to be effective. View factor correlations result in reliable and reasonably accurate predictions. For a solid volume fraction of 0.45, the root mean squared errors of particle-particle and particle-wall view factors are $2.7\times10^{-4}$ and 0.021 with corresponding training data in the ranges of 0–0.08 and 0–0.5 respectively. To scale these correlations for large number of particles, restricting shading detection up to 5 nearest neighbors is demonstrated to be an effective strategy to balance prediction accuracy with computational efficiency. With thousands of particles, the computational cost of proposed view factor correlations with thresholding of 5 shading particles is about 100 times faster than serial Monte Carlo ray tracing simulations for a solid volume fraction of 0.45.

**Nomenclature**

| | |
|---|---|
| $a, b, c$ | fit-function coefficients for the effect of shading function |
| $C$ | correction factor for particle-wall |
| $d$ | distance or diameter, mm |
| $f$ | function |
| $F$ | view factor |
| $H$ | height of modeling domain, mm |
| $k$ | number of shading particles |
| $l$ | Euclidean norm |
| $L$ | length of modeling domain, mm |
| $M$ | the maximum power of input variable |
| $N$ | number |
| $O$ | complexity |
| $P$ | probability |
| $r$ | line vector |
| $S$ | shading factor for particle-particle |
| $W$ | width of modeling domain, mm |
| $x$ | pertaining to x-coordinate or input variable |
| $X$ | input matrix |
| $y$ | pertaining to y-coordinate or output variable |
| $Y$ | output matrix |
| $z$ | pertaining to z-coordinate |

*Greek symbols*

| | |
|---|---|
| $\alpha$ | viewing angle magnitude, ° |
| $\beta$ | coefficient for polynomial functions |
| $\gamma$ | ratio or normalized value |
| $\phi$ | volume fraction |

*Superscripts*

| | |
|---|---|
| $m$ | power of input variable |
| $T$ | transpose |
| $*$ | pertinent to a dimensionless value |
| $-$ | pertinent to a mean value |
| $\sim$ | pertinent to a predicted value |
| $\rightarrow$ | vector |

*Subscripts*

| | |
|---|---|
| $c$ | pertinent to a critical value |
| $i$ | the index of launching surface, including particles |
| $im_1j$ | lines connecting particle centers of $i, m_1$ and $i, j$ |
| $j$ | the index of intercepting surface, including particles and six walls |
| $m_1$ | nearest shading particle |
| $m_2$ | the second nearest shading particle |
| max | the maximum value |
| min | the minimum value |
| $n$ | index of input variable |
| $p$ | pertinent to a particle or summation index of scaling factor |
| $pp$ | particle-particle |
| $pp,0$ | particle-particle without shading effect |
| $pp,k$ | particle-particle with $k$ number of shading particles |
| $pw$ | particle-wall |
| $pw,0$ | particle-wall without shading effect |
| $pw,k$ | particle-wall with $k$ number of shading particles |
| $q$ | summation index of scaling factor |
| *s* | solid |
| *t* | tangential or training dataset |
| *v* | validation dataset |
| *w* | wall |
| ⊥ | normal direction |

*Abbreviations*

| | |
|---|---|
| MCRT | Monte Carlo ray tracing |
| MRE | mean relative error |
| RMSE | root mean squared error |
| $R^2$ | R squared value or the coefficient of determination |

# 1. Introduction

Heat transfer in particulate media has important fundamental and technological applications, and thermal radiation becomes a dominant mode of heat transfer at high temperatures and in evacuated environments [1]–[6]. Particulate media enhances heat and mass transfer between the solid and fluid phases such as in packed beds [7]–[10], foams and fibers [11]–[14], granular flows and fluidized beds [3], [15]–[18]. Such media are radiatively participating, as thermal radiation can be absorbed, emitted, and scattered within their volume. Many models have been developed to predict radiative transport in static porous and dispersed media [7], [19]–[21]. However, they can be somewhat limited to evaluate radiative transport in dynamically changing participating media involving flowing particles and chemical reactions. Flow regimes dictate spatial distributions of particles and therefore its effective radiative properties [22], [23], and chemical transformations can affect material composition and therefore its properties [24], [25]. Such dynamic participating media find important applications as heat-transfer and thermochemical materials in concentrated solar power plants [2], [3], [26], and in reactors for drying, catalysis and gasification for fuel production applications [4]–[6], [27]. Our study focuses on the determination of radiative view factor correlations as a function of particle spatial locations and plane wall dimensions in an ensemble to facilitate discrete radiative flux calculations for flowing particles.

Radiative fluxes in the energy transport equation can be evaluated using deterministic and probabilistic approaches. The former numerically/computationally solves the continuum radiative transport equation with *a priori* knowledge of its effective radiative properties, including the extinction coefficient, scattering albedo, and phase function [28]. Many approximations have been developed to simplify the governing radiative transport equation specific to its application [29], [30]. The diffusion and $P_1$ approximations are well-suited to quantify radiative fluxes in optically thick media and in regions precluded from large gradients in energy densities [31]–[33]. The discrete ordinates approximation assumes finite angular directions for radiative intensity transport and provides the advantage of seamlessly integrating with finite volume solvers for other transport phenomena (mass, momentum, energy, species) [34]–[37]. However, due to the discrete nature of the angular approximations, "ray effects" can arise due to spurious, large spatial oscillations in the radiative energy density [38]. This effect has been overcome by increasing the number of angular directions, averaging over angular quadratures with different reference frame orientations [39] and by adding artificial scattering of radiative intensity [40], but with the drawback of increase in compute/memory requirements. Moreover, convergence and stability need to be examined with the number of angular directions in addition to the mesh density for the discrete ordinates method [37]. For packed beds of large particles, the diffusion approximation [41]–[43], and discrete ordinate method [8], [34] are commonly used techniques to model radiative transport. However, beyond the constraints and limitations already discussed, these techniques rely on inputs for the effective radiative properties obtained from models and/or measurements.

Probabilistic approach can be used to launch and trace many rays or photon bundles for absorption, scattering and emission events in a participating medium [44], [45]. This approach has been used to determine effective radiative properties for participating media [46], [47], to directly model radiative fluxes [48], [49], and to also compute radiative view factors [49]–[51]. A high degree of accuracy can be achieved with enough rays (typically ~$10^6$) being tracked and by using physics-based probability density functions. However, this approach will be especially limiting for the cases of flowing particles due to large computational time and memory requirements [50][52]. Therefore, radiative transport in particulate flows have been modeled using ray tracing by either analyzing the system at selected snapshots of time [53] or by considering very small volume fractions of particles (< 0.006) to perform flow-radiation coupling using ray tracing [54].

Other than deterministic and stochastic predictions, data-driven modeling for radiative transport in participating media is also gaining traction [55]–[60]. Wu et al. [57] and Tausendschön et al. [58] have developed neural network models to obtain particle-particle and particle-wall view factors based on distance normalized by diameter. While the former did not generalize their results for different solid volume fractions, the latter developed geometry-based correlations which were however shown to be sensitive to the solid volume fraction of the training dataset. Therefore, these correlations are better suited for moderately dense particle beds (solid volume fraction of 0.2–0.4). Johnson et al. have developed view factor correlations based on particle positions, and applied it to model radiative heat transfer in the gravity-driven flow through a channel [59], [60]. However, this study did not consider the effect of shading by neighboring particles and applied average view factor values as a function of distance between particles. This approximation can lead to large deviations in predicted view factors for high solid volume fractions, where shading effects on view factor values becomes significant. Even though shading effects were considered to predict particle-particle view factor in the study by Feng and Han's [61], these results were not further interpreted to develop position based correlations. Overall, most of the existing data-driven view factor predictions either are applicable to a specific morphology and/or preclude the effects of shading by neighboring particles, especially for particle-wall view factor calculations. Additionally, physically interpretable view factor predictions that also account for shading effects are missing.

Motivated by current knowledge gaps, the primary objective of this study is to determine particle-particle and particle-wall view factor correlations as a function of spatial locations of the particles and wall surfaces for an ensemble of large particles in a packed bed, while considering shading effects. Data-driven modeling based on multivariate linear regression is used to obtain the governing correlations. Training and validation datasets for the particle-particle and particle-wall view factors are obtained from Monte Carlo ray tracing (MCRT) simulations. Ray tracing is performed for a random packed bed of spherical particles with solid volume fractions ranging from 0.016 to 0.45. To isolate effects of shading by neighboring particles, pairs of particles with varying number of particles in between them were modeled using ray tracing, which informs data-driven shading factor predictions. Compared to prior work with similar scope, our study is the first to obtain view factor correlations accurately and efficiently while factoring in the effects of shading. Even though the training dataset of view factors were obtained from a static bed, because the correlations developed only depend on spatial locations/positions, particle size and plane wall dimensions, they can be extended to compute pairwise view factors even for flowing particles, where particle spatial locations can be dynamically updated with time. Additionally, our correlations present a more computationally lightweight approach to compute radiative view factors compared to collision-based ray tracing evaluations.

## 2. Theory and Modeling Approach

A packed bed with monodisperse and randomly distributed spherical particles was computationally generated to obtain surface-surface view factors (Fig. 1). Particles with a fixed diameter of 0.4 mm were placed inside a large enough domain with dimensions (4 mm × 4 mm × 6 mm) that are at least ten times larger than the particle size in any direction. Particle sizes are considered to be large enough compared to the characteristic wavelengths for thermal radiation, such that geometric optics is applicable to model radiation [62], [63]. The absolute size of the particles modeled, and the modeling domain will not affect the predictions, as these correlations were obtained as a function of non-dimensional parameters. The coordinates for particle centers within the bed were randomly sampled from a uniform distribution of spatial locations within the domain, with the constraint of no overlaps between any pair of particles generated. If an overlap was detected, the coordinates of particle position were regenerated. While particle centers are always located inside the cuboid, some fraction of the surface area of particles can lie outside the bounding surfaces (Fig. 1). Five distinct solid volume fractions were considered, $\phi_s = 0.016$, 0.068, 0.12, 0.28, and 0.45, by varying the number of particles, $N_p$, inside the domain (Eq. (1)).

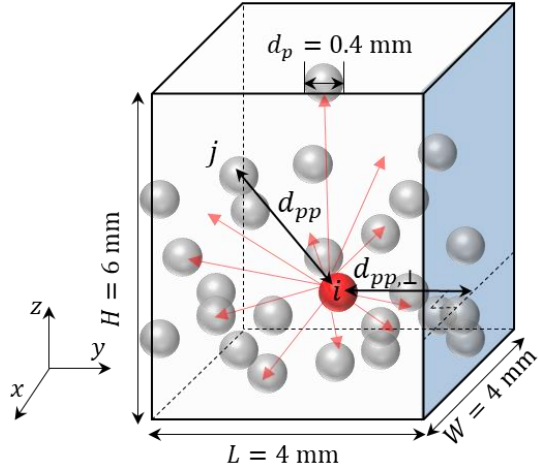

**Fig. 1**. Schematic of packed bed modeling domain, in a size of 4 mm × 4 mm × 6 mm ( length × width × height), with spherical particles of diameter, $d_p = 0.4$ mm, for MCRT simulations to predict particle-particle and particle-wall view factors. Millions of rays are launched from target particle $i$ (red) and traced for intersections with other surfaces including every other particle $j$ (grey) plus six walls. The inter distance between particles $i$ and $j$ and normal distance between particle $i$ and the right wall (blue shaded) are annotated as $d_{pp}$ and $d_{pp,\perp}$ (mm).

$$\phi_s = \frac{N_p \times \frac{4}{3} \times \pi \times \left(\frac{d_p}{2}\right)^3}{L \times W \times H} \quad (1)$$

Solid volume fraction range selected in this study is representative of dilute to moderately dense packing of particles. For chemical catalysis applications, packed beds with solid volume fractions in the range of 0.35–0.65 have been extensively used [64]–[66]; in particle receivers for concentrated solar power applications solid volume fractions of 0.01–0.34 are expected for gravity-driven flow of sand-like particles [67], [68], and solid volume fractions of 0.01–0.15 are commonly encountered in fluidized bed combustors [69]–[72].

### 2.1. Monte Carlo Ray Tracing Simulations for View Factor Predictions

Radiative view factors, also referred to as exchange factors, are geometric parameters that quantify the fraction of radiative energy leaving one surface that is intercepted by another surface [73]. Collision-based Monte Carlo ray tracing (MCRT) simulations were performed to evaluate the diffuse view factor, (Eq. (2)) between pairs of particles, and between particle and wall surfaces.

$$F_{ij} = \frac{\text{Total number of rays intercepted by surface } j}{\text{Total number of rays launched diffusely from particle surface } i} \qquad (2)$$

A statistically large number of rays, up to $10^7$, are launched from every particle and traced for intersections with other particles and bounding wall surfaces. The launch coordinates were sampled to be uniformly distributed on the particle surface (Section A1). Particle surfaces were assumed to be diffuse, which is a reasonable choice for unpolished material surfaces with roughness. The polar and the azimuthal angle for the launched rays were sampled from physics-informed cumulative distribution functions to diffusely emit rays from the particle surface. All surfaces, including particles and the walls, were modeled to be perfectly absorptive to determine the view factors [73]. Therefore, when a ray intersects any surface, its tracking is complete and followed by the launch of a new ray. Rays were launched from every particle surface and tracked for intersections with every other particles and wall surfaces in the domain. For particle-particle view factor, $i$ and $j$ are both particles, and therefore $F_{ij}$ is denoted as $F_{pp}$; similarly, the particle-wall view factor is denoted as $F_{pw}$. For $N_p$ particles, we compute $N_p(N_p + 6)$ view factors; for the largest solid volume fraction modeled, this amounts to about 1.6 million (~$1289^2$) total view factors computed. A flowchart has been provided in Section A1 to detail the algorithm for computing view factors using MCRT simulations.

An in-house C++ MCRT code, previously developed by Li et al. [22], was adapted and modified to compute view factors. Statistical convergence was ensured by launching and tracking a large enough number of rays that yielded minimal changes in the predicted view factors with increase in the number of rays launched. Changes in view factors were quantified by computing the $l_2$ norm with respect to results obtained for the case with $10^7$ rays for a solid volume fraction of $\phi_s = 0.28$ with 800 particles. The $l_2$ norm of 0.0179, 0.0058 and 0.0020 were obtained for $10^4$, $10^5$, and $10^6$ rays respectively, indicating statistical convergence for $10^6$ rays. Simulation results for view factors were validated by, (a) checking for the criteria of view factor summation, self-viewing, superposition and reciprocity [73] (Section A2), and (b) by comparison with analytical solutions [74] of a pair of spherical particles (Section 3.1.). With $10^6$ rays the summation criterion is satisfied perfectly, whereas reciprocity criterion is satisfied with 4.7% error. Ray tracing predictions are within 1.3% of the analytical solution for a pair of particles.

The ray tracing simulations were compiled with Microsoft Visual Studio Community 2019 and performed on an Intel® Core™ i7-9700 processor (3.00GHz, 32GB). For the largest solid volume fraction modeled, $\phi_s = 0.45$, the view factor computations from ray tracing took about 14 hours of wall-clock time without parallelization.

## 2.2. Data-driven Modeling for View Factor Correlations

Fig. 2 shows the flowchart of the algorithm in this study. Particle-particle, $F_{pp}$, and particle-wall, $F_{pw}$, view factors obtained from MCRT simulations were used as inputs to train and validate data-driven models. Data-driven correlations and analytical solutions are first obtained to predict the maximum particle-particle and particle-wall view factor values in the absence of any shading effects. Next, informed by ray tracing simulation data, we develop geometry-based relationships to correct for shading effect as a function of a shading factor based on $k$ shading particles and viewing angle, $S_{pp,k}$, for particle-particle view factor. Similarly, data-driven correlations are developed to compute a correction factor, $C_{pw}$, which scales the net contributions of shading from particles present between a particle and a wall surface. Predicted data were compared with the validation dataset from the MCRT simulations to optimize best-fit parameters and functions in the view factor correlations obtained. Roughly 80% of view factors from ray tracing simulations were randomly selected and used for training, and the balance was used for validation.

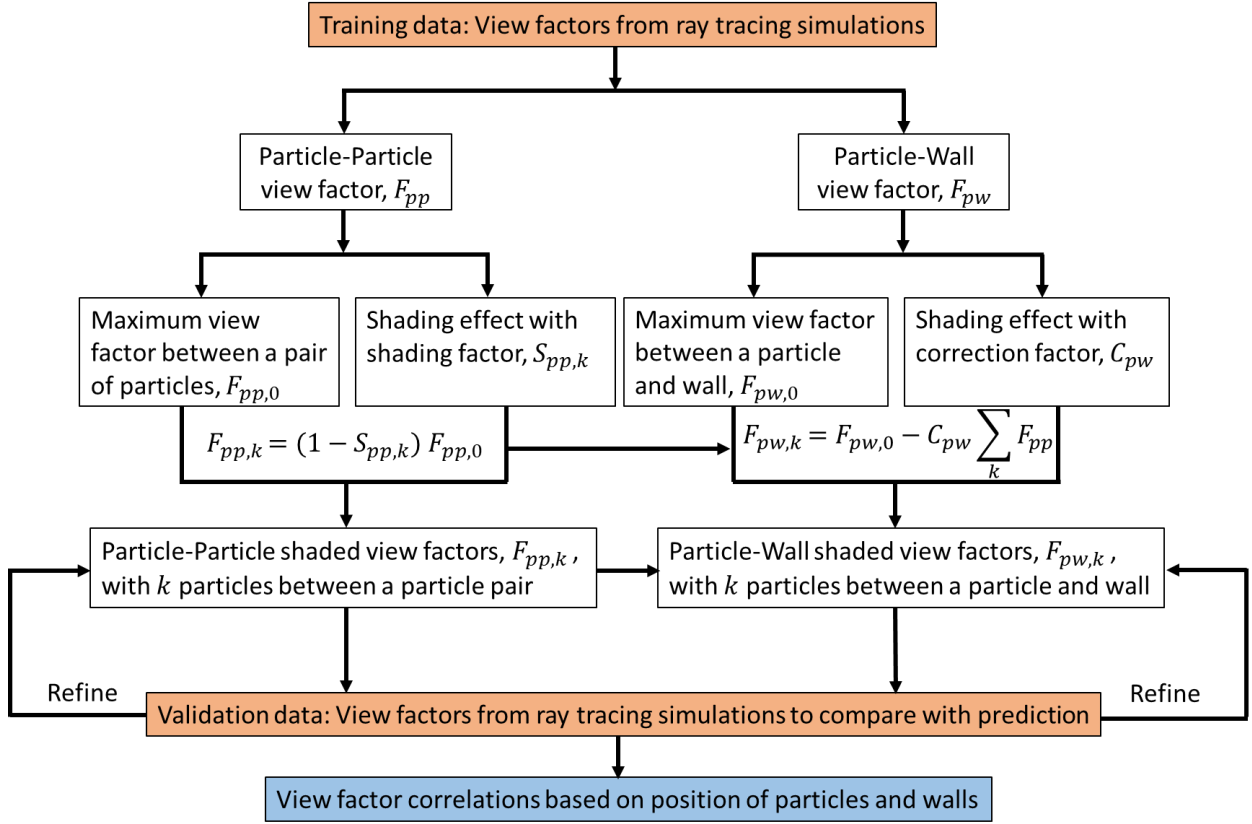

**Fig. 2.** Flowchart depicting the use of ray tracing simulations to provide training and validation datasets to develop data-driven/ reduced-order correlations for particle-particle and particle-wall view factors as a function of the spatial location of particles and wall surfaces and while accounting for shading effects.

Polynomial functions were considered in Eq. (3) to predict the maximum particle-particle view factor without shading, $y_n = F_{pp,0}$ and to determine the correction factor that scales the extent of shading by particles in particle-wall view factors (Fig. 1), i.e., $y_n = C_{pw}$, with regression coefficients, $\beta_m$, and feature variables, $x_n$, with varying power $m$. The power is ranging from an integer value of $m_{\min}$ to a maximum value of $M$, and $N_t$ is the size of the training dataset.

$$y_n = \sum_m \beta_m x_n^m \qquad \begin{array}{l} n = 1, 2, \ldots, N_t \\ m = m_{\min}, m_{\min} + 1, \ldots, M \end{array} \qquad (3)$$

The description of the feature variable depends on what view factor is being predicted. For particle-particle view factors, it is the ratio of the inter-particle distance between pairs of particles, $d_{pp}$, to the diameter of the particles, $d_p$ (Eq. (4)). However, for the correction factor predictions, the feature variable is the non-dimensional normal distance, $d_{pw}^*$ between a select particle and a wall surface, as in Eq. (5).

$$x_n = d^*_{pp} = \frac{d_{pp}}{d_p}; p = 1 - N_p \tag{4}$$

$$x_n = d^*_{pw} = \frac{d_{pw,\perp}}{d_p}; p = 1 - N_p, w = 1 - 6 \tag{5}$$

In matrix form, Eq. (3) can be rearranged as Eq. (6a), and the expanded expression is shown in Eq. (6b).

$$X\beta = Y \tag{6a}$$

$$X = \begin{bmatrix} x_1^{m_{min}} & x_1^{m_{min}+1} & \cdots & x_1^{M-1} & x_1^M \\ x_2^{m_{min}} & x_2^{m_{min}+1} & \cdots & x_2^{M-1} & x_2^M \\ \cdots & \cdots & \cdots & \cdots & \cdots \\ x_{N_t}^{m_{min}} & x_{N_t}^{m_{min}+1} & \cdots & x_{N_t}^{M-1} & x_{N_t}^M \end{bmatrix}, \quad \beta = \begin{bmatrix} \beta_{m_{min}} \\ \beta_{m_{min}+1} \\ \cdots \\ \beta_{M-1} \\ \beta_M \end{bmatrix}, \quad Y = \begin{bmatrix} y_1 \\ y_2 \\ \cdots \\ y_{N_t} \end{bmatrix} \tag{6b}$$

The $\beta$ values in Eq. (6a) were obtained by solving the linear equation using matrix inversion (Eq. (7)),

$$\beta = (X^T X)^{-1} X^T Y \tag{7}$$

where, $X^T$ is the transpose of the matrix $X$. Using the closed-form solution for $\beta$ works well in this case because of the relatively small size of the datasets considered.

The best-fit functions are obtained by optimizing the $\beta$ values to minimize the root mean square error (RMSE) in Eq. (8), by considering the differences between the data, $y_n$, from MCRT simulations and predictions, $\widetilde{y_n}$, from different regression models; RMSE is averaged over the total number of validation datasets, $N_v$. The coefficient of determination, $R^2$, provides a measure of the quality of the fit by comparing the deviation of model predictions with the variance obtained based on a mean value, $\overline{y_n}$, of the validation dataset (Eq. (9)). Additionally, to quantify relative variation between actual and predicted values, the mean relative error (MRE) is calculated using $y_n$ from MCRT data and $\widetilde{y_n}$ from regression model predictions (Eq. (10)). Since $y_n$ can be an extremely small value or even 0, this relative error is only computed for a subset of the data, $N'$, where $y_n > 10^{-5}$.

$$\text{RMSE} = \sqrt{\frac{1}{N_v} \sum_{n=1}^{N_v} (y_n - \widetilde{y_n})^2} \tag{8}$$

$$R^2 = 1 - \frac{\sum_{n=1}^{N_v}(y_n - \widetilde{y_n})^2}{\sum_{n=1}^{N_v}(y_n - \overline{y_n})^2} \tag{9}$$

$$\text{MRE} = \frac{1}{N'} \sum_{\substack{n=1 \\ y_n > 10^{-5}}}^{N'} \frac{|(y_n - \widetilde{y_n})|}{y_n} \tag{10}$$

Multivariate linear regression codes to obtain optimal $\beta$ values were developed and implemented in MATLAB R2019b using Intel® Core™ i7-9700 processor (3.00GHz 32GB).

### 2.2.1. Particle-Particle View Factors

Closed form expressions has been reported to determine the maximum particle-particle view factor, $F_{pp,0}$, for a pair of spherical particles without any shading particles for large distances between particles, and a lookup table/discrete numerical values exist for small distance regimes as shown in Eq. (11) [74].

$$F_{pp,0} = \begin{cases} \text{Numerical values}, & d_{pp}^* < 2.5 \\ \frac{1}{2}\left(1 - \left(1 - \frac{1}{4d_{pp}^{*2}}\right)^{\frac{1}{2}}\right), & d_{pp}^* \geq 2.5 \end{cases} \quad (11)$$

Using MCRT results for a pair of spherical particles, we obtained a closed-form expression for a wider range of inter-particle separation distances of $1 < d_{pp}^* \leq 20$.

MCRT simulations were performed to independently predict the effects of shading by 1 and 2 particles present in between a pair of particles (Fig. 3). Results from these cases were used to also quantify shading effects in the presence of more than 2 particles. As will be discussed in the results section, shading due to larger than 10 shading particles is most likely to result in particle-particle view factors that are very close to 0 (Section 3.1.). Consider the geometry set up when there is one particle, $m_1$, between a pair of particles $i$ and $j$ (Fig. 3 (a)). View factor between particles $i$ and $j$ was computed in the presence of particle $m_1$, and as a function of viewing angles between particles. Ray tracing simulations were performed with $10^6$ rays launched from the surface of particle $i$ and traced for intersections with $j$. To probe the influences of the viewing angle, the angular position of particle $j$ was varied relative to the positions of $i$ and $m_1$ (Fig. 3 (a)). For this calculation, while the coordinates of the centers of particles $i$ and $m_1$ were fixed, the position of particle $j$ was varied as a function of the viewing angle, $\alpha_{im_1 j}$, for a selected distance, $d_{ij}$. This angle formed between the line vectors connecting the centers of particle $i$ and $m_1$, and particle $i$ and $j$, and its calculation is shown in Fig. 3. For a viewing angle of $\alpha_{im_1 j} = 0°$, all three particles are along the same line vector and there will be full shading of particle $j$ by $m_1$. Because this angle is computed as a magnitude, it also accounts for particle $j$ being rotated counterclockwise from the line vector connecting $i$ and $m_1$. View factors were computed as a function of the viewing angle, $\alpha_{im_1 j}$, from these calculations. Even though the distances between pairs of particles were ($d_{im_1} = 0.7$ mm, $d_{ij} = 1.4$ mm) fixed, as will be shown in the results, it doesn't impact the generality of the proposed prediction algorithm for monodisperse ensembles of particles.

The same approach was also extended to compute shading effects from the presence of two particles, $m_1$ and $m_2$, between a pair of particles $i$ and $j$ (Fig. 3 (b)). In this case, the relative angular positions of particles $m_2$ and $j$ were varied with respect to fixed particle centers for $i$ and $m_1$. The distances, $d_{im_1}, d_{im_2}$ and $d_{ij}$ were 0.7, 1.4 and 2.7 mm respectively for these calculations. Correspondingly, we predict the dependency of particle-particle view factor between $i$ and $j$ as a function of two viewing angles, $\alpha_{im_1 j}$ and $\alpha_{im_2 j}$.

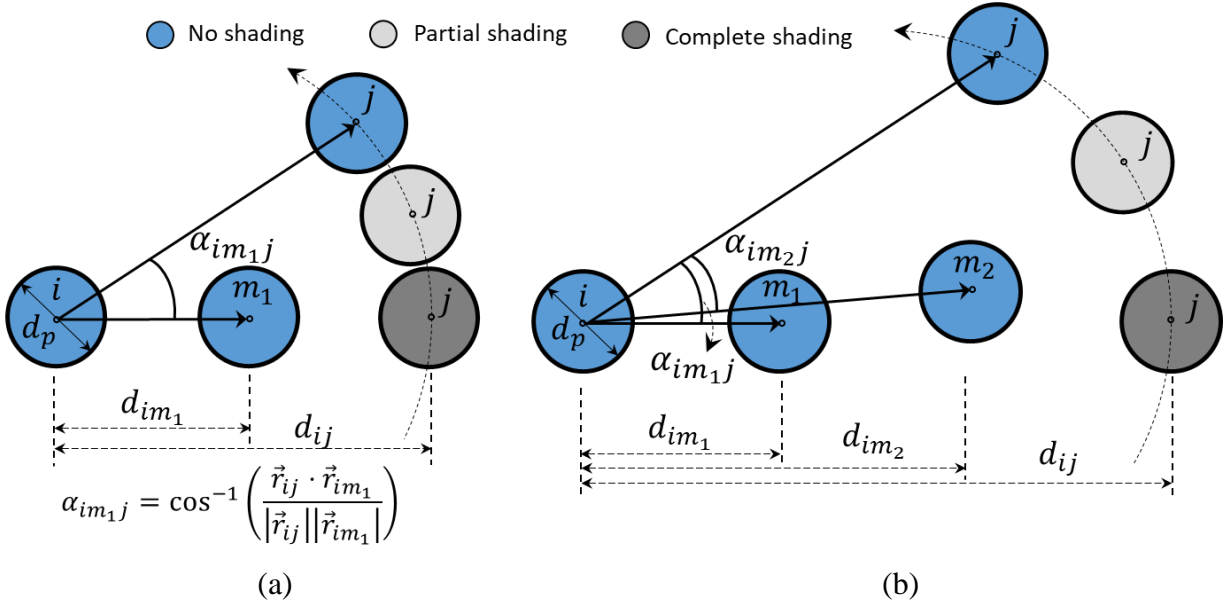

**Fig. 3.** Model set up to compute shaded view factors with (a) one particle, $m_1$ and (b) two particles $m_1$, $m_2$, present between a pair of particles $i$ and $j$. Different relative positions of particle $j$ with respect to particles $i$, $m_1$ and $m_2$ can lead to cases of no, partial and complete shading. Viewing angle $\alpha_{im_1 j}$ quantifies the magnitude of the angle between the line vectors, $\vec{r}_{im_1}$ and $\vec{r}_{ij}$, that connecting the centers of particles $i$ and $m_1$, with $i$ and $j$; for the sake of illustration this angle is shown when particle $j$ is at no shading position; similarly, viewing angle $\alpha_{im_2 j}$ is defined as well.

Results from these calculations were used to inform two complimentary quantities — $\gamma_{pp,k}$ and $S_{pp,k}$ (Eqs. (12a) and (12b)).

$$\gamma_{pp,k} = \frac{F_{pp,k}}{F_{pp,0}}, k = 0, 1, 2, \dots, k_{\max} \tag{12a}$$

$$S_{pp,k} = 1 - \gamma_{pp,k} \tag{12b}$$

The normalized view factor, $\gamma_{pp,k}$, is the ratio of the shaded particle-particle view factor with $k$ particles present between any pair of particles to the maximum particle-particle view factor, and a shading factor, $S_{pp,k}$, that determines the extent of normalized deviation of shaded view factors from their maximum values. Both the normalized view factor and the shading factor lie between 0–1. When the shading factor value is $S_{pp} = 0$ (equivalent to $\gamma_{pp} = 1$), it indicates no shading, whereas a value of $S_{pp} = 1$ (equivalent to $\gamma_{pp} = 0$) indicates complete shading. As will be shown in results (Section 3.1.), the scaling factor is dependent on viewing angle, and whether a particle is a shading particle ($S_{pp,1} > 0$) is determined by the comparison between viewing angle and critical viewing angle. Specifically, any particle, $m_k$, present between a pair of particles ($i$ and $j$) will cast a shade on particle $j$ as viewed from particle $i$, when the viewing angle $\alpha_{im_k j}$ is smaller than the critical angle, $\alpha_{c,im_k}$, subtended by particle $i$ on particle $m_k$ (Fig. 7). This criterion is also used to identify the likelihood of number of shading particles as a function of solid volume fraction (Section A6). This calculation is useful for finding a threshold number of shading particles after which further accounting for shading doesn't make a big influence on the calculated view factors.

The number of shading particles between any particle pairs can range from 0, when there is no shading, to a maximum of 50 for $\phi_s = 0.45$.

## 2.2.2. Particle-Wall View Factors

Particle-wall view factors are corrected for shading effects from the analytical view factor for one particle viewing a plane wall, $F_{pw,0}$, in Eq. (13a) obtained as a function of dimensionless distances (Eq. (13b)), where, $x_p$, $y_p$, $z_p$ are the spatial coordinates of the particle, $H$ and $W$ are the height and width of the right wall in Fig. 1; for other walls, the appropriate values are used for the height and the width [73].

$$F_{pw,0} = f\left(\frac{d_{pw,\perp}}{x_p}, \frac{d_{pw,\perp}}{z_p}\right) + f\left(\frac{d_{pw,\perp}}{x_p}, \frac{d_{pw,\perp}}{H-z_p}\right) + f\left(\frac{d_{pw,\perp}}{W-x_p}, \frac{d_{pw,\perp}}{z_p}\right) + f\left(\frac{d_{pw,\perp}}{W-x_p}, \frac{d_{pw,\perp}}{H-z_p}\right) \quad (13a)$$

$$f(d_1^*, d_2^*) = \frac{1}{4\pi} \tan^{-1}\left(\left(d_1^{*2} + d_2^{*2} + d_1^{*2}d_2^{*2}\right)^{-\frac{1}{2}}\right) \quad (13b)$$

In the presence of additional particles, the view factor between a particle and a wall surface should be less than that predicted by Eq. (13b). This is because rays leaving the particle of interest can be obstructed by particles in between itself and the wall surface. The number of intermediate particles can be as large as 1287 for $\phi_s = 0.45$ for particle-wall view factor calculations, contrasting a maximum of 50 for particle-particle shading considerations for the same solid volume fraction. Therefore, we propose corrections due to shading effects from these intermediate particles by introducing a particle-wall correction factor, $C_{pw}$ (Eq. (14)), which ranges from 0 to 1.

$$F_{pw,k} = F_{pw,0} - C_{pw} \sum_{k=1}^{k \leq 450} F_{pp'_k} \quad (14)$$

In Eq. (14)), this correction factor scales the sum of the particle-particle view factors between the particle of interest, $p$, and the intermediate particles, $p'_k$, up to a maximum of 450, to capture the net shading effect from all the relevant intermediate particles. Beyond 450 intermediate particles, the particle-wall view factors become small ($< 5\%$ of maximum value) and considering shading effects from more particles does not make the prediction any more accurate – RMSE changes by less than 1% for $k = 450$ compared to $k = 1287$. Even though this cut-off number of intermediate particles is dictated by the largest solid volume fraction, $\phi_s = 0.45$, it is not expected to vary significantly for $\phi_s > 0.45$.

MCRT simulation results for the shaded particle-wall view factors, $F_{pw,k}$ and our data-driven model predictions for particle-particle view factor, $F_{pp'_k}$ were used to obtain correlations for the correction factor, $C_{pw}$, as a function of the dimensionless normal distance between the particle and the wall, $d^*_{pw}$ (Eq. (5)). Using this dataset, data-driven models were trained by considering different hypothesis functions for $C_{pw}$ (Section A7) and determining best-fit values by applying regression technique (Eqs. (3), (5)–(10)). From training data, it is observed that with an increase in the dimensionless particle-wall distance, the sum of the intermediate particle-particle view factors increases, and the correction factor decreases. Therefore, at small values of $d^*_{pw}$, $C_{pw}$ approaches a value of 1 with fewer intermediate particles, which results in the shaded particle-wall view factors approaching the analytical solution, i.e., $F_{pw,k} \to F_{pw,0}$ (Eq. (14)). At large $d^*_{pw}$, $C_{pw}$ approaches about 0.1, because many intermediate particles at least partially cast their shade on the wall, and this increases the deviation between the shaded particle-wall view factor and the analytical solution.

## 3. Results and Discussion

### 3.1. Particle-Particle View Factors

Fig. 4 shows the MCRT predictions for particle-particle view factors, $F_{pp}$, as a function of the dimensionless distance, $d_{pp}^*$ (Eq. (4)) for the solid volume fractions modeled, $\phi_s = 0.016$–$0.45$. For comparison, the maximum view factor between a pair of particles, which is independent of $\phi_s$, is also included. At any solid volume fraction, the particle-particle view factors decrease rapidly with increasing dimensionless distance. This is driven by both a decrease in solid angle with increased separation and an increase in shading by neighboring particles. For small distances ($d_{pp}^* \leq 1.3$) all the predicted view factors deviate from the maximum view factor by at most 10% for any solid volume fraction and is attributed to the low likelihood of shading effects. However, for larger particle separation ($d_{pp}^* > 1.3$) there is a more significant influence of the solid volume fraction on the predicted view factors. With increase in $\phi_s$, there is an increase in the spread of view factors that lie between 0 and the maximum view factor value (inset in Fig. 4). For dimensionless distances in the range of $2 \leq d_{pp}^* \leq 4$, only a small fraction of the predicted data lies below the maximum view factor value for $\phi_s = 0.016$, while the spread becomes substantially larger for $\phi_s = 0.45$. At large distances, $8 \leq d_{pp}^* \leq 10$, $\phi_s = 0.016$ has a larger spread in the data compared to $\phi_s = 0.45$. This is because shading effects are strong enough for the larger solid volume fraction that all the view factor values become small ($< 2.5$e-4). Even though shading effects are not as significant for low solid volume fractions, it is still important enough to yield a spread in the data. Results in Fig. 4 were further interpreted to determine that the threshold distance between particles, beyond which shading effects become important decreases with increase in the solid volume fraction. For instance, all the predicted view factors are within 10% deviation from the maximum view factor at $d_{pp}^* \leq 2.7$ for $\phi_s = 0.016$. Contrastingly, this threshold distance is reduced by more than half to $d_{pp}^* \leq 1.3$ for $\phi_s = 0.45$. For large particle-particle distances at $d_{pp}^* \geq 12$, even the maximum view factor values become small ($< 4.3$e-4) at any solid volume fraction. This value is less than 1% of the maximum view factor value computed at $d_{pp}^* = 1.3$. Therefore, for large distances, particle-particle view factors with shading effects can be reasonably approximated as 0.

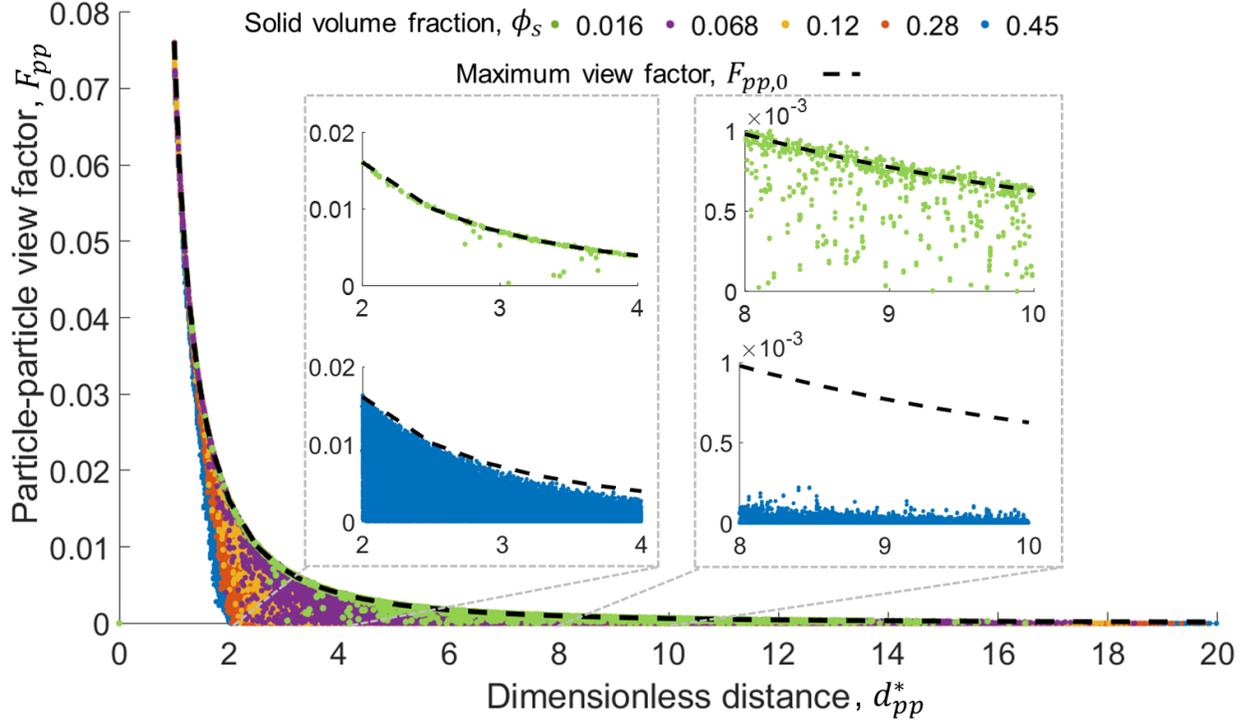

**Fig. 4.** Particle-particle view factors, $F_{pp}$, from Monte Carlo ray tracing (MCRT) simulations performed for a random packed bed with monodisperse and diffuse spheres with a size of 0.4 mm (Fig. 1) with respect to dimensionless distances, $d_{pp}^*$, for solid volume fractions, $\phi_s = 0.016$, 0.068, 0.12, 0.28, and 0.45. Subplots show the detailed data for $\phi_s = 0.016$ and 0.45 for dimensionless distances of 2–4 and 8–10.

To further probe the effects of $\phi_s$ and $d_{pp}^*$ on shading, Fig. 5 shows the probability distributions of the normalized view factors, $\gamma_{pp}$, at selected distances of $d_{pp}^* = 2.525$, 5.025, 7.525. At every $d_{pp}^*$, normalized view factors are computed for dimensionless distances that are within $\pm 0.025$ deviation around the listed mean values. For $\phi_s = 0.016$ and small distances (Fig. 5(a)), the likelihood of having view factor values larger than 80% of the maximum value is 1. However, for $\phi_s = 0.45$, the likelihood decreases as the normalized view factor value increases from 0 to 1 (Fig. 5(d)). With increasing distance between particles, stronger shading effects result in a more dispersed normalized view factor distribution for $\phi_s = 0.016$ (Fig. 5(b), (c)) and leads to high likelihood of small values of $\gamma_{pp} < 0.1$ for $\phi_s = 0.45$. (Fig. 5(e), (f)). These results reinforce the necessity to correct for shading effects in the determination of particle-particle view factors and highlight how dominant these effects are for the larger solid volume fractions.

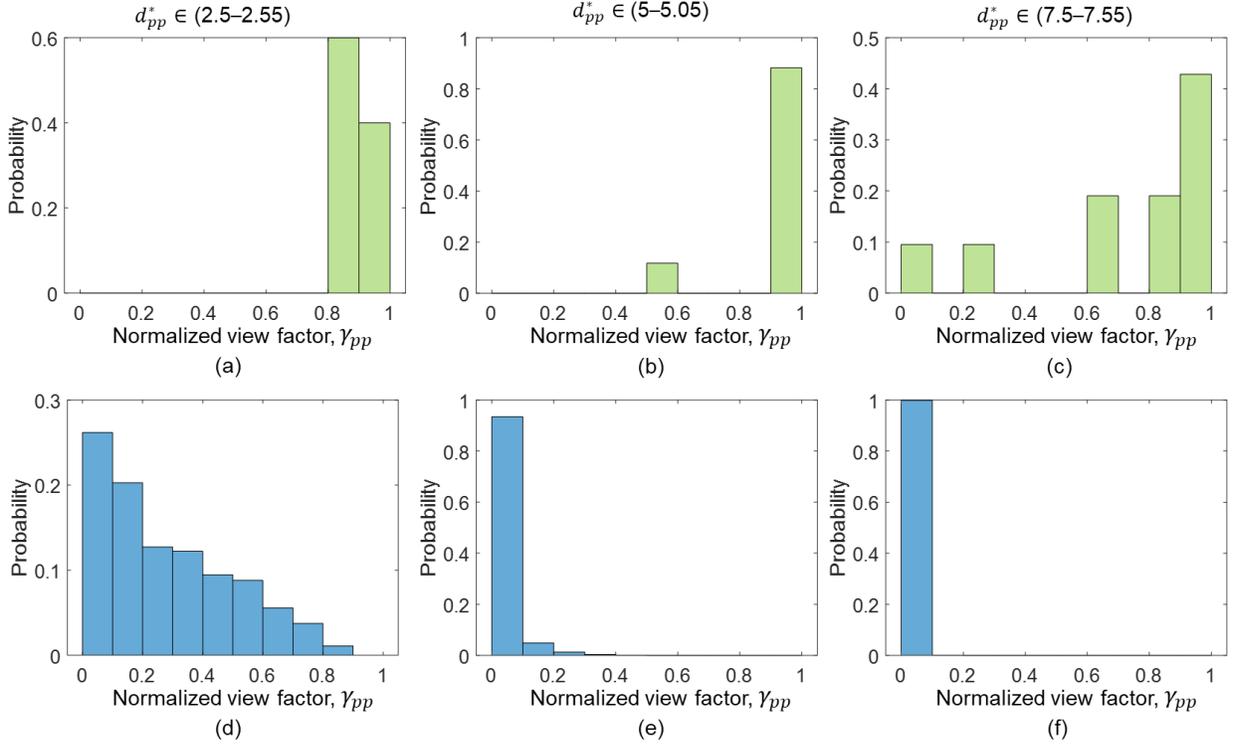

**Fig. 5.** Probability distributions of normalized particle-particle view factors, $\gamma_{pp}$, at different dimensionless distance of (a), (d) 2.5–2.55, (b), (e) 5–5.05, (c), (f) 7.5–7.55 for solid volume fractions of (a)–(c) $\phi_s = 0.016$ and (d)–(f) $\phi_s = 0.45$. Particle-particle view factors were obtained from Monte Carlo ray tracing (MCRT) simulations performed for a random packed bed with monodisperse and diffuse spheres with a size of 0.4 mm.

Table 1 shows the root mean-squared error, RMSE (Eq. (8)), between the predicted and the validation datasets from MCRT simulations for the various hypothesis functions tested to predict the maximum value of the particle-particle view factor. The best-fit function that yielded reasonably low RMSE values (7e-5) combined with large $R^2 = 0.99998$ comprises five feature variables dependent on dimensionless distance, $d^{*m}$, with $m$ varying from -4 to 0 ($f_7$). Compared to this function, a fourth-order polynomial function ($f_1$), results in RMSE values that are larger by about two orders-of-magnitude. The importance of the presence of the $1/d^{*2}$ term in the view factor correlation is illustrated by the substantially smaller value for RMSE with functions, $f_3$, compared to the functions that only included the $1/d^*$ and $1/d^{*3}$ dependence — $f_2$ and $f_4$ respectively. This result is physically reasonable and consistent with the intrinsic solid angle definition that varies proportional to $1/d^{*2}$. When more negative $m$ terms in $d^{*m}$ ($1/d^{*5}$, $1/d^{*6}$, etc.) are included in functions $f_8$–$f_{10}$, the RMSE value only marginally decreases from the prediction in $f_7$, but $R^2$ is no longer changing, likely due to overfitting the data. Therefore, $f_7$ is selected as the best-fit function as it achieves accurate predictions and good fit quality with fewer fitting parameters compared to the other functions. Detailed plot is shown in Section A3 for selected functions. The maximum view factors are from 0.075 to 1.56e-4 for $d^*$ in 1–20 and assured to be non-negative.

**Table 1**: Ten hypothesis functions to predict the maximum particle-particle view factors (Eq. (3)) with coefficients obtained from multivariate linear regression with feature variables based on dimensionless distance, $d_{pp}^*$, which is shortened as $d^*$ for brevity. The corresponding RMSE and $R^2$ are shown.

| Function # | Hypothesis Functions | RMSE | $R^2$ |
|---|---|---|---|
| $f_1$ | $y = 8.8\text{e-}2 - 4.4\text{e-}2d^* + 7.2\text{e-}3d^{*2} - 4.6\text{e-}4d^{*3} + 1\text{e-}5d^{*4}$ | 5.4e-3 | 0.90507 |
| $f_2$ | $y = -7\text{e-}3 + 6.2\text{e-}2/d^*$ | 3.9e-3 | 0.95279 |
| $f_3$ | $y = -4.7\text{e-}4 + 7.1\text{e-}2/d^{*2}$ | 7.1e-4 | 0.99836 |
| $f_4$ | $y = 1.9\text{e-}3 + 8\text{e-}2/d^{*3}$ | 2.0e-3 | 0.98733 |
| $f_5$ | $y = 1.1\text{e-}3 - 1.3\text{e-}2/d^* + 8.5\text{e-}2/d^{*2}$ | 5.1e-4 | 0.99914 |
| $f_6$ | $y = -4.6\text{e-}4 + 8\text{e-}3/d^* + 3.2\text{e-}2/d^{*2} + 3.6\text{e-}2/d^{*3}$ | 1.5e-4 | 0.99992 |
| $f_7$ | $y = 3.0\text{e-}4 - 5.6\text{e-}3/d^* + 9.4\text{e-}2/d^{*2} - 6.3\text{e-}2/d^{*3} + 4.9\text{e-}2/d^{*4}$ | 7.0e-5 | 0.99998 |
| $f_8$ | $y = -1.8\text{e-}4 + 4.4\text{e-}3/d^* + 2.9\text{e-}2/d^{*2} + 1\text{e-}1/d^{*3} - 1.3\text{e-}1/d^{*4} + 7.1\text{e-}2/d^{*5}$ | 6.9e-5 | 0.99998 |
| $f_9$ | $y = 1.3\text{e-}4 - 4.1\text{e-}3/d^* + 1\text{e-}1/d^{*2} - 1.9\text{e-}1/d^{*3} + 4\text{e-}1/d^{*4} - 3.9\text{e-}1/d^{*5} + 1.5\text{e-}1/d^{*6}$ | 6.9e-5 | 0.99998 |
| $f_{10}$ | $y = -9.7\text{e-}5 + 3.2\text{e-}3/d^* + 2.5\text{e-}2/d^{*2} + 2.1\text{e-}1/d^{*3} - 6.1\text{e-}1/d^{*4} + 9.9\text{e-}1/d^{*5} - 8.1\text{e-}1/d^{*6} + 2.7\text{e-}1/d^{*7}$ | 6.3e-5 | 0.99998 |

Fig. 6 shows the dependence of the normalized particle-particle view factor on the viewing angle, $\alpha_{im_1j}$ in a 3-particle system with particle $m_1$ in between particles $i$ and $j$ (Fig. 3). For $\alpha_{im_1j} = 0$, there is complete shading as all 3 particles are along the same line, and this results in $\gamma_{pp,1} = 0$. However, there is a non-zero view factor value for all other viewing angles because all particles are modeled as diffuse emitters. This is because for any viewing angle $\alpha_{im_1j} > 0$, some fraction of the rays leaving particle $i$ will still intercept particle $j$. As the viewing angle increases, the view factor initially rapidly increases, after which the rate of increase slows down until it attains the maximum value, where particle $j$ is no longer shaded by particle $m_1$. The oscillation in the predicted maximum view factor value is an artifact of the stochastic nature of MCRT simulations. For fixed distances between pairs of particles ($d_{im_1}$ and $d_{ij}$ in Fig. 3(a)), a larger viewing angle reduces shading effects, and beyond a critical viewing angle, $\alpha_{c,im_1}$, the predicted view factor approaches the maximum value within 2%. This critical angle is approximately twice the tangential angle, $\alpha_{t,im_1}$, between particles $i$ and $m_1$ for any combination of $d_{im_1}$ and $d_p$ values (Eq. (15)),

$$\alpha_{c,im_1} \cong 2\alpha_{t,im_1} = 2\sin^{-1}\left(\frac{0.5d_p}{d_{im_1}}\right) \qquad (15)$$

The tangential angle in Eq. (15), is the magnitude of the angle between the tangent from the center of particle $i$ to particle $m_1$, and the line connecting the center of particle $i$ and $m_1$. When the viewing angle is twice the tangential angle, geometrically, particle $j$ is almost completely out of the shadow-zone cast by particle $m_1$ as viewed from particle $i$. Therefore, view factor values beyond this critical angle approach the maximum view factor value.

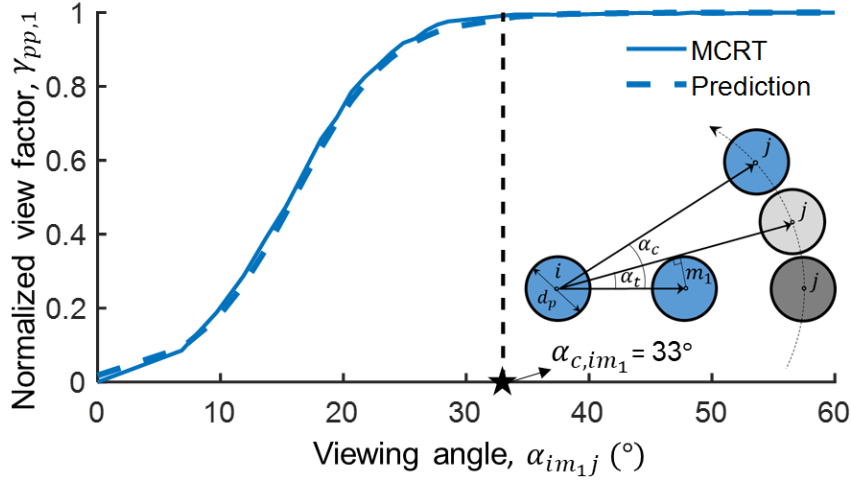

**Fig. 6.** Normalized particle-particle view factor with one shading particle, $\gamma_{pp,1}$, as a function of viewing angle $\alpha_{im_1 j}$, from Monte Carlo ray tracing (MCRT) simulations (solid line) and the best-fit prediction (dashed line) using a sigmoid function in Eq. (16a) with $d_{ij} = 1.4$ mm, $d_{im_1} = 0.7$ mm, and $d_p = 0.4$ mm in Fig. 3. In the inset is a schematic of the 3-particle system with critical, $\alpha_c$, and tangential, $\alpha_t$ angles annotated. The black star and vertical line correspond to the critical viewing angle between particles $i$ and $m_1$ beyond which shading effect by particle $m_1$ is not significant.

A modified sigmoid function (Eq. (16a)) with three coefficients (Eq. (16b)) fits the functional dependence of the normalized view factor due to shading by 1 particle, $\gamma_{pp,1}$, on the viewing angle, $\alpha_{im_1 j}$. Data from these predictions when validated against MCRT simulations yield RMSE = 0.017 and $R^2 = 0.9933$.

$$\gamma_{pp,1} = \frac{F_{pp,1}}{F_{pp,0}} = \frac{a}{1 + \exp\bigl(-c(\alpha_{im_1 j} - b)\bigr)} \tag{16a}$$

$$a = 1; \; b = \alpha_t = \sin^{-1}\left(\frac{0.5 d_p}{d_{im_1}}\right); \; c = \frac{4}{\alpha_t} \tag{16b}$$

The numerator in a sigmoid function in Eq. (16a) is the limiting value attained for large values of the independent variable. The predicted view factor asymptotes to the maximum view factor, $F_{pp,0}$, for large viewing angles, which results in the normalized view factor and therefore the numerator, $a$, in Eq. (16a) being equal to 1 (Fig. 6). From Eq. (16a), it is evident that when $\alpha_{im_1 j} = b$, the shaded view factor attains 50% of the maximum value. From results in Fig. 6, it is observed that when $\alpha_{im_1 j} = \alpha_t$, the predicted view factor is nearly 50% of the maximum view factor, and this dictates the value of $b$ (Eq. (16b)). The coefficient $c$ in the dominator is determined by making the predicted view factor value attain about 98% of maximum value, which occurs when $\alpha_{im_1 j} = 2\alpha_t$. It cannot be 100% due to the inherent nature of sigmoid function, which attains the exact maximum value as $\alpha_{im_1 j} \to \infty$. The 95% confidence intervals for the fitted coefficients in Eq. (16b) are listed in Section A4. Although the result shown in Fig. 6 is for a specific combination of distances between particles, $d_{im_1}$ and $d_{ij}$, the deduced correlation in Eq. (16a) is applicable more generally for any monodisperse distribution of particles, as fitted coefficients in Eq. (16b) are non-dimensionalizing particle-particle separation distances with particle size. While coefficients $a, b, c$ in Eq. (16a) are drawn from geometry-based parameters, they are within 5% deviation from those obtained via curve-fitting a sigmoid function in MATLAB.

Fig. 7 applies a similar approach as in Fig. 6 to show the effects of shading in a 4-particle system with two shading particles $m_1$ and $m_2$ (Fig. 3). The normalized shaded particle-particle view factors, $\gamma_{pp,2}$, are computed as a function of two viewing angles $\alpha_{im_1 j}$ and $\alpha_{im_2 j}$. For selected particle-particle distances ($d_{im_1} = 0.7, d_{im_2} = 1.4, d_{ij} = 2.7$ mm), the critical viewing angles are $\alpha_{c,im_1} = 33°$ and $\alpha_{c,im_2} = 16°$ respectively. For any position of particle $m_2$, when $\alpha_{im_1 j} = 0°$, and equivalently, for any position of particle $m_1$, when $\alpha_{im_2 j} = 0°$, the respective particle triplets — ($i, m_1, j$) and ($i, m_2, j$) are along the same line. This results in complete shading at the boundaries in Fig. 7 with $\gamma_{pp,2} = 0$.

Four regions can be identified in Fig. 7(a) based on the relative values of $\alpha_{im_1 j}$ and $\alpha_{im_2 j}$ with respect to the respective critical angles, $\alpha_{c,im_1}$ and $\alpha_{c,im_2}$, and these regions are visually illustrated with sample particle locations in Fig. 7(b). In region I, both viewing angles are smaller than their respective critical angles, and therefore, particle $j$ is shaded by both $m_1$ and $m_2$ (Fig. 7(b)). Therefore, in this region, both viewing angles can influence the value of $F_{pp,2}$ (Fig. 7(a)). In regions II ($\alpha_{im_1 j} > \alpha_{c,im_1}$ and $\alpha_{im_2 j} < \alpha_{c,im_2}$) and III ($\alpha_{im_1 j} < \alpha_{c,im_1}$ and $\alpha_{im_2 j} > \alpha_{c,im_2}$), shading effects arise from only one particle, either particle $m_2$ or particle $m_1$ respectively for Regions II and III (Fig. 7(b)). Distinct from Region I, the shaded view factor in these two regions depend on only one viewing angle (Fig. 7(a)). Therefore, the trends are like the 3-particle system in Fig. 6, where an increase in viewing angle leads to an increase in view factor and decrease of shading effect. In Region IV, when both viewing angles are larger than their respective critical angles, there is no effect of shading by either $m_1$ or $m_2$ (Fig. 7(b)), and $\gamma_{pp,2} \approx 1$, i.e., the shaded view factor nearly equals the maximum view factor (Fig. 7(a)). Oscillations in the numerical values of $\gamma_{pp,2}$ in Regions IV are due to the stochastic nature of MCRT simulations.

From Fig. 7, we determine that the shading effects and therefore shading factor, $S_{pp,2}$, because of two particles will be a non-linear function of shading factors by particle $m_1$ ($S_{1-m_1}$) or $m_2$ ($S_{1-m_2}$) alone. $S_{pp,2}$ is shortened as $S_2$ for brevity. This prediction is especially important in Region I where both particles affect shading. If the overall scaling factor $S_2$ is larger than 1, it is reset as 1 to ensure that predictions for view factors are non-negative. Different functional forms were tested to predict $S_2$ as a function of $S_{1-m_1}$ and $S_{1-m_2}$ and presented in Section A5. The best-fit function presented in Eq. (17) yielded a RMSE of 0.032 and included linear additions of the individual shading factors, $S_{1-m_1}$ and $S_{1-m_2}$, and a product term to compensate the overestimation of shading by considering independent contributions by two particles.

$$S_2 = \text{Min}(1, S_{1-m_1} + S_{1-m_2} - S_{1-m_1} S_{1-m_2}) \tag{17}$$

Eq. (17) has been generalized in Eq. (18) for $k$ particles present between particles $i$ and $j$,

$$S_k = \text{Min}\left(1, \sum_{p=1}^{k} S_{1-m_p} - \sum_{p=1}^{k-1} \sum_{q>p}^{k} S_{1-m_p} S_{1-m_q}\right); k = 1\text{–}50 \tag{18}$$

where, $k$ ranges from 1 to a maximum of 50 for $\phi_s = 0.45$, and the product term that is summed over $k$ particles will have $\frac{k(k-1)}{2}$ terms. Table A4 illustrates the expanded equations for $k = 1\text{–}5$.

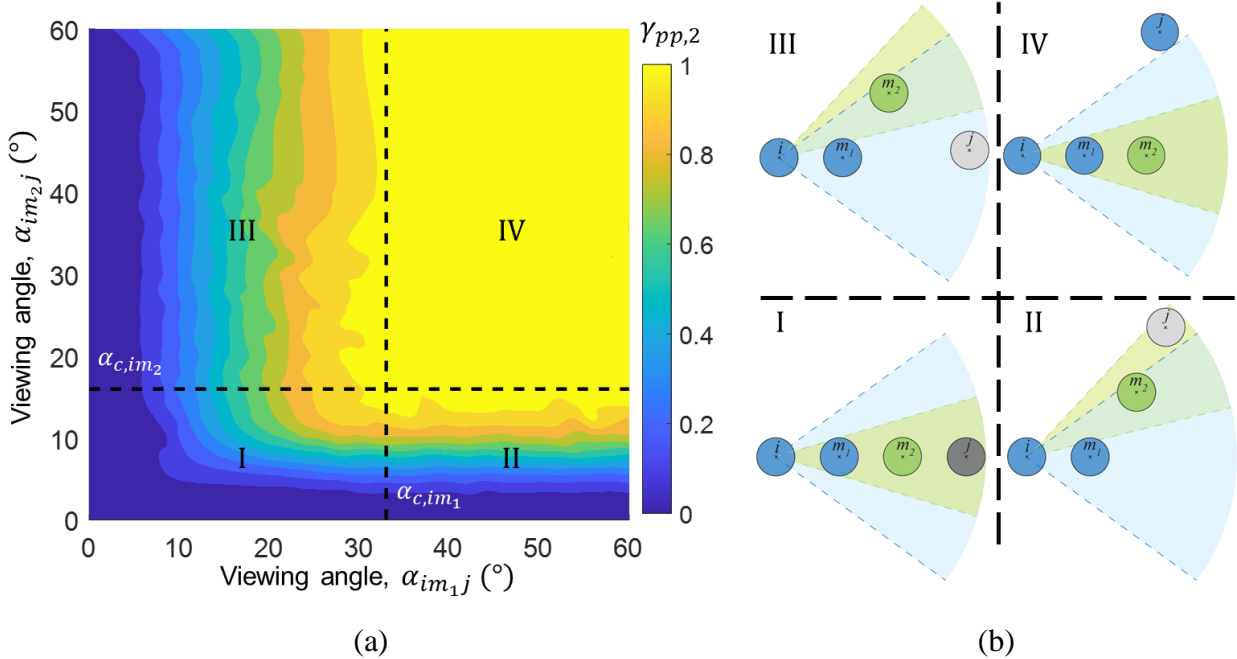

**Fig. 7.** Effects of viewing angles — $\alpha_{im_1 j}$ and $\alpha_{im_2 j}$ — in a 4-particle system with two shading particles on (a) normalized particle-particle view factor, $\gamma_{pp,2}$, with the critical viewing angles for particle $m_1$ and $m_2$ — $\alpha_{c,im_1}$ and $\alpha_{c,im_2}$ represented as dashed lines leading to (b) four regions, I: $m_1$ and $m_2$ cast shadows, II: $m_2$ cast shadow, III: $m_1$ cast shadow and IV: no shading by neither $m_1$ nor $m_2$; shadow zones of particles $m_1$ and $m_2$ are illustrated in blue and green based on expected tangential angles of $m_1$ and $m_2$. MCRT simulation data for this 4-particle system was obtained with $d_{im_1} = 0.7$ mm, $d_{im_2} = 1.4$ mm, $d_{ij} = 2.7$ mm, and $d_p = 0.4$ mm.

Fig. 8 shows that the proposed extension in Eq. (18) results in reasonable comparisons with MCRT predictions of shaded particle-particle view factors; $\phi_s = 0.016$ (Fig. 8(a)) and $\phi_s = 0.45$ (Fig. 8(b)) are shown for conciseness. Datasets are categorized also by the number of shading particles, $k$, present between the designated pair of particles, which was computed based on the viewing angle, $\alpha_{im_k j}$, and critical angle, $\alpha_{c,im_k}$ (Fig. 7). For every value of the number of shading particles, a maximum of 20 data points are randomly selected and shown in Fig. 8 to avoid visual overcrowding; for $\phi_s = 0.016$ and 5 shading particles, there are only 2 data points to plot. Expectedly, at equivalent values of the dimensionless distance, the number of shading particles between any pair of particles is larger for the larger solid volume fraction. The RMSE values reported in the tables are however calculated for all pairwise view factors predicted. The RMSE values are 8.3e-5 and 2.7e-4 for $\phi_s = 0.016$ and 0.45 respectively, which amounts to a MRE of 8.7% and 85%. With the increase in solid volume fraction, prediction errors increase due to increased errors in accounting for shading from neighboring particles and the large magnitudes of relative errors stem from a high density of small view factor values ($< 1e-4$). Therefore, deviations of small values get amplified in relative error estimations. The RMSE values exhibit a somewhat non-monotonic variation with the number of shading particles, especially for $\phi_s = 0.45$, where these errors are smaller when more than 5 particles cause shading compared to when shading occurs due to at most 5 particles. This is because the RMSE values are biased by the large density of view factors with very small magnitudes ($< $ 1e-4) when $k > 5$ for the high solid volume fraction case.

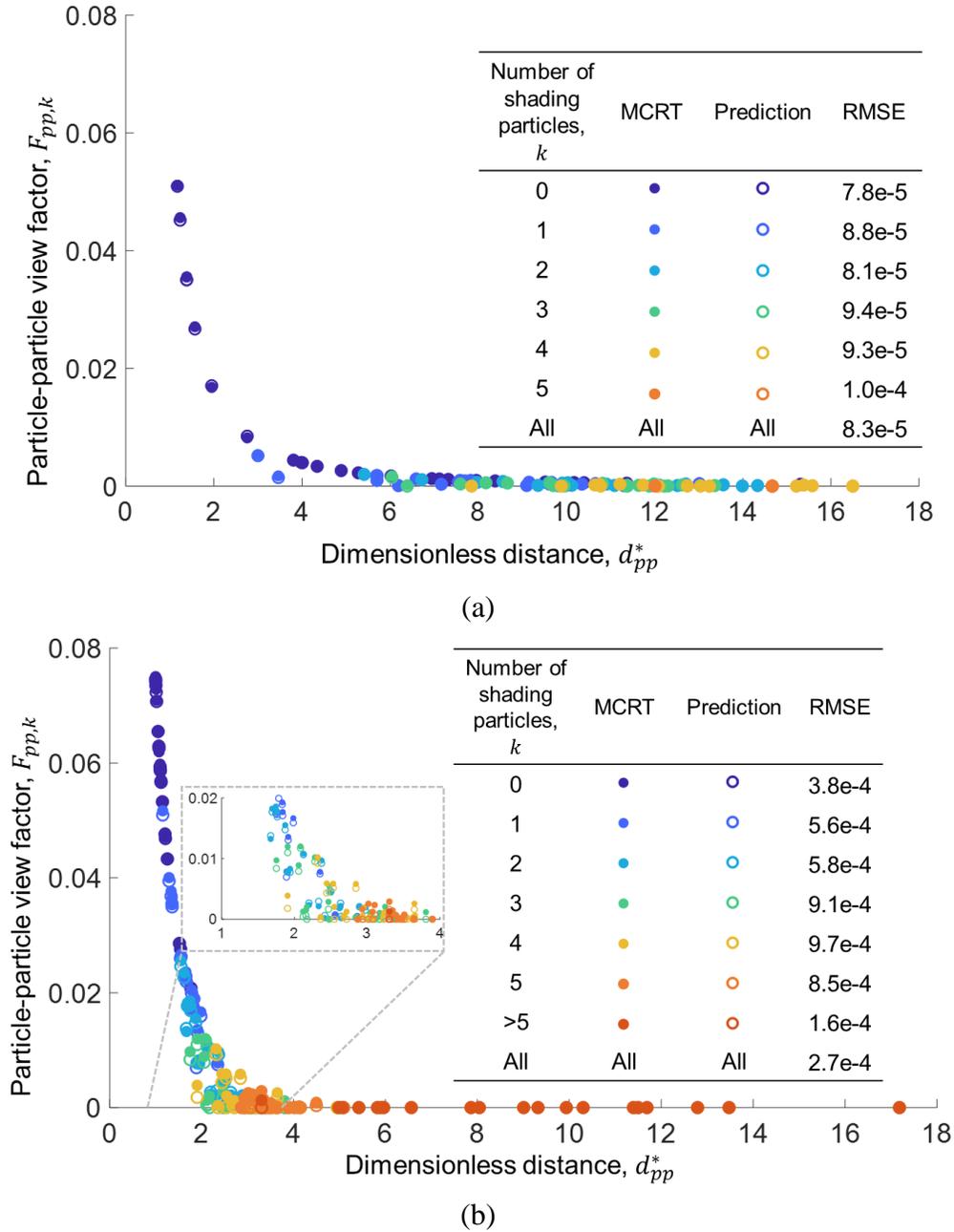

**Fig. 8.** Particle-particle view factor predictions from $f_7$ in Table 1 and Eq. (4), (12a), (15)–(18) compared with Monte Carlo ray tracing (MCRT) simulations with different colors representing different number of shading particles, $k$, for (a) $\phi_s = 0.016$ and (b) $\phi_s = 0.45$ with root mean squared error (RMSE) reported in the inset tables as a function of $k$; $k$ value of *All* refers to all the view factors with different $k$. MCRT simulations for particle-particle view factors were performed for a random packed bed with monodisperse and diffuse spheres with a size of 0.4 mm.

## 3.2. Particle-Wall View Factor

Fig. 9 shows the training data and predictions for the particle-wall correction factor, $C_{pw}$ as a function of the dimensionless particle-wall normal distance, $d^*_{pw}$, (Eq. (5)) for all 6 walls; training data was determined from Eq. (14). It is observed that $C_{pw}$ decreases with increase $d^*_{pw}$, and the best-fit function and the parameters are influenced by the solid volume fraction. Therefore, distinct best-fit functions are reported for low ($\phi_s = 0.016, 0.068, 0.12$, Fig. 9(a)) and high ($\phi_s = 0.28, 0.45$, Fig. 9(b)) solid volume fractions in Eqs. (19a) and (19b) respectively.

$$C_{pw} = 0.67 - 0.059 d^*_{pw} + 0.0016 {d^*_{pw}}^2, \phi_s \leq 0.12 \tag{19a}$$

$$C_{pw} = 0.90 - 0.098 d^*_{pw} + 0.0032 {d^*_{pw}}^2, 0.12 < \phi_s \leq 0.45 \tag{19b}$$

The fit quality is mediocre with RMSE errors up to 0.0078 and the $R^2$ values being less than 0.9 and attributed to the spread in the training data at any $d^*_{pw}$. This spread can stem from: (i) scatter in the predictions for $F_{pw,0}$ (Eq. (13)) due to the sensitivity to lateral positioning of particles; (ii) variations in the contributions of intermediate particles to shading, and (iii) $C_{pw}$ being predicted solely as a function of $d^*_{pw}$ (Eq. (14)), when in actuality it might depend on more than one feature. For the sake of simplicity, this study retains the last assumption, and as will be shown (Fig. 10), the fit quality for $C_{pw}$ doesn't negatively impact the particle-wall view factor predictions, $F_{pw,k}$. Other prediction functions that were tested for $C_{pw}$ included higher order polynomial functions and sigmoid functions with only $d^*_{pw}$ as the feature variable (Section A7), but they did not result in any substantial improvements to the quality of fits compared to the quadratic functions considered.

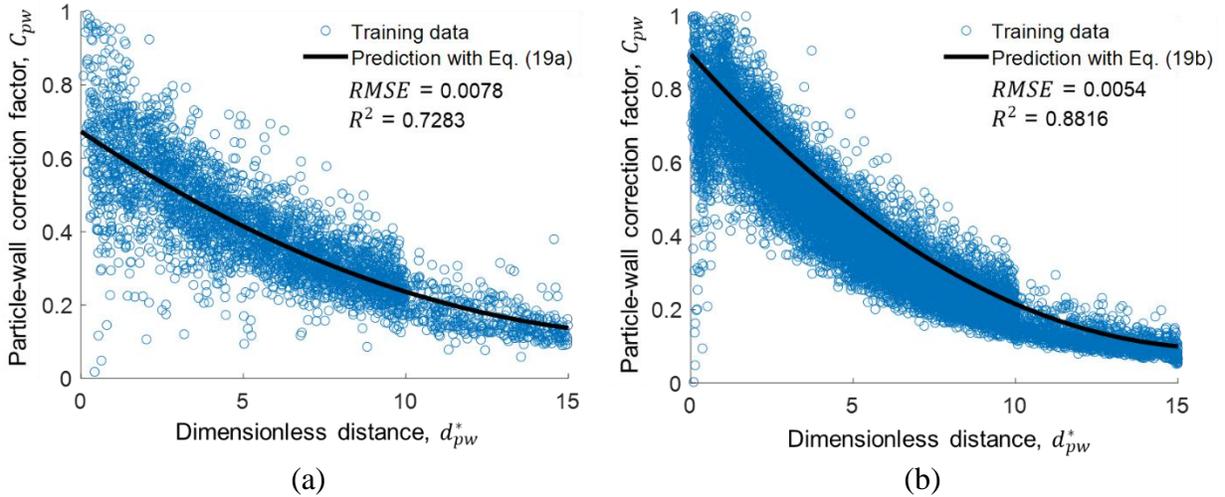

(a)      (b)

**Fig. 9.** Comparison of training data (blue) and prediction (black) from Eq. (19) for the particle-wall correction factor, $C_{pw}$, with respect to the dimensionless particle-wall normal distance, $d^*_{pw}$, RMSE and $R^2$ values are shown in the plot for (a) low ($\phi_s = 0.016, 0.068, 0.12$) and (b) high solid volume fraction ($\phi_s = 0.28, 0.45$). The density of the data in both plots is comparatively smaller for $d^*_{pw} > 10$ because of the rectangular shape of modeling domain. All training data for $C_{pw}$ were deduced from Monte Carlo ray tracing (MCRT) simulation results for particle-wall view factors with monodisperse and diffuse spheres with a size of 0.4 mm.

Fig. 10 depicts the predicted values for particle-wall view factors, $F_{pw,k}$, using $C_{pw}$ obtained from Fig. 9 and using Eq. (14). These predictions are compared against MCRT simulation data for particle-wall view factors, and analytical estimates for view factors without shading, $F_{pw,0}$, (Eq. (13a)) for $\phi_s = 0.016$ and $0.45$. View factors are shown only for particles with the right wall, but the results for the other wall surfaces follow suit (insets in Figs. 10(a) and (b)). Despite the average quality of fits for $C_{pw}$ (Fig. 9), the predictions match well with ray tracing simulations with RMSE values of 0.0089 and 0.021 for $\phi_s = 0.016$ and $0.45$ respectively. These RMSE values translate to MRE of 7.1% and 87% respectively. For the lowest solid volume fraction modeled, even the analytical solution matches the predictions from MCRT simulations quite well (RMSE = 0.021) because the particle-particle shading effects are not as pronounced. However, for the high solid volume fraction, shading effects become significant and the analytical results substantially overpredict the view factors at any distance leading to an order-of-magnitude larger RMSE values compared to predictions from our correlations (Fig. 10(b)). The analytical particle-wall view factor especially exhibits a large spread when the normal distance between the particle and the wall is lesser than 1.2 mm, corresponding to $d_{pw}^* < 3$. This is because, when a particle is near a wall, its lateral positioning can severely impact the extent to which it views the wall. Contrastingly, for the same $d_{pw}^*$ values, MCRT simulation results that account for shading effects indicate lesser scatter in the data, due to the significant obstructions caused by the presence of intermediate particles. This trend is well-captured by our predictions for the shaded view factors, $F_{pw,k}$, from Eq. (14).

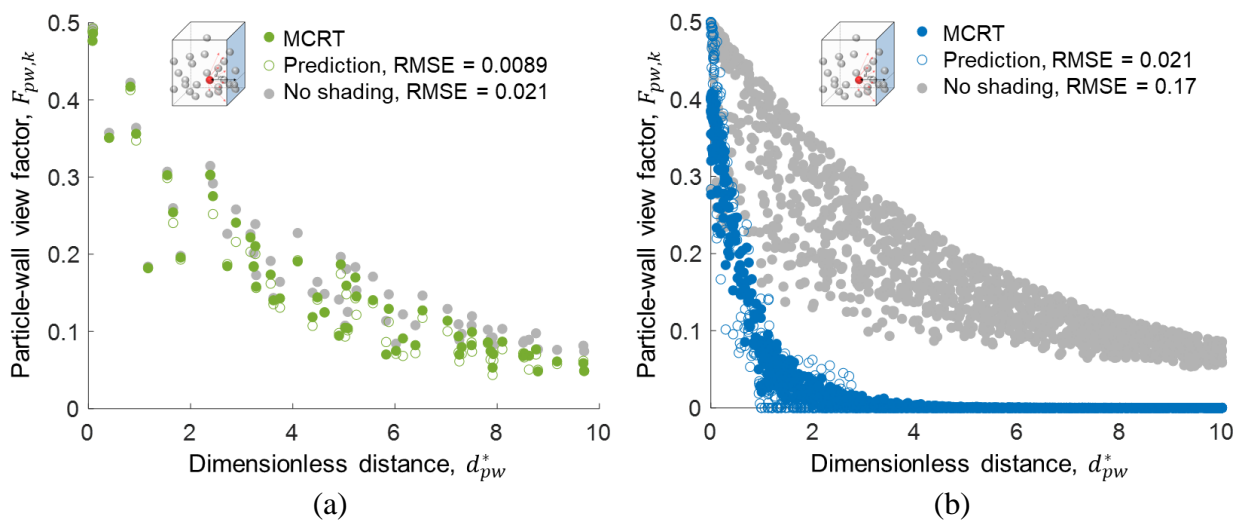

**Fig. 10.** Comparison of view factors between particles and the right wall surface (highlighted in the inset) between Monte Carlo ray tracing (MCRT) simulation data, predictions from Eq. (14) for shaded particle-wall view factor, $F_{pw,k}$, and analytical data for particle-wall view factor without shading, $F_{pw,0}$, for (a) $\phi_s = 0.016$ and (b) $\phi_s = 0.45$ as a function of the dimensionless normal distance, $d_{pw}^*$; root mean squared error (RMSE) is reported for our predictions and analytical solutions against MCRT simulations. All MCRT simulations for particle-wall view factors were obtained from a random packed bed with monodisperse and diffuse spheres with a size of 0.4 mm.

### 3.3. Computational Accuracy and Efficiency for Particle-Particle View Factors

Computing particle-particle view factors can pose formidable challenges, to especially account for shading by particle neighbors, when the number of particles becomes large. Particle-wall view factors are also dependent on, and limited by, the calculation of particle-particle view factors (Eq. (14)). Therefore, Fig. 11 assesses different strategies for thresholding based on the number of shading particles to compute shaded particle-particle view factors, $F_{pp,k}$ (Eq. (18)). Thresholding values imply that the prediction is performed up to a variable number, $k$, of shading particles, and beyond this number, i.e., $F_{pp,k+1} = 0$ in Eq. (18). Comparisons of different thresholding approaches are made based on accuracy and computational cost for the predictions.

The prediction errors — RMSE and MRE from view factor correlations developed in this work ($f_7$ in Table 1, Eqs. (4), (12), (15)–(18) for $F_{pp,k}$ and Eqs. (5), (13)–(14), (19) for $F_{pw,k}$) are shown in Fig. 11(a), (b) as a function of the threshold number of shading particles. For all solid volume fractions, the steepest reduction in the prediction errors occurs when the thresholding value for the shading particles increases from 0 to 5; a threshold value of 0 implies that no shading corrections are made and the maximum particle-particle view factor, $F_{pp,0}$ ($f_7$ in Table 1) is used as such. For comparison, we also show the errors for the *no-prediction* case, where all particle-particle view factor values are set to be 0, i.e., $F_{pp,k} = 0$ for all $k$. For $\phi_s = 0.016$, the maximum number of shading particles is 5, and therefore the data do not extend beyond this value. For $\phi_s = 0.45$, which can have shading by up to 50 particles, the errors level off when accounting for only 5 nearest shading particles. Results from a thresholding value of 10 nearest particles lead to the same prediction errors as accounting for all possible shading particles. This is consistent with the findings from the probability distributions of the particle-particle view factors as a function of the number of shading particles (Section A6), which indicate that there is very low probability of view factors being even 10% of the maximum view factor with more than 10 shading particles. Additionally, the probability of having 10 shading particles is also low for all solid volume fractions. Therefore, accurate predictions can be made for view factors from our correlations by accounting for shading effects by 5, and no more than 10, nearest neighboring particles for $\phi_s$ in the range of 0.016–0.45. The RMSE and MRE increase as the solid volume fraction increases because of the increased errors in predicting shading corrections. MRE for the large solid volume fractions is biased by the high density of particle-particle view factors with small magnitudes (< 1e-4). Error distribution plot (Section A8) as a function of view factor reveals that relative error drops off drastically to less than 10% for view factors larger than 0.02.

Fig. 11(c) shows computational time as a function of the number of particles and equivalently, the solid volume fraction, and its dependency on thresholding values of 5 for the number of shading particles. Compared to serially executed ray tracing simulations, correlations for view factors developed in this study is more time-efficient only when thresholding is applied to detect shading particles. The algorithm complexity scales as the square of the number of particles, $O(N_p^2)$, where $N_p$ is the number of particles, for both MCRT simulations and the correlation-based approach with thresholding. When all the shading particles are considered, the complexity becomes $O(N_p^3)$, which can especially become penalizing when $N_p$ becomes large. Even with the same complexity, the reason why the MCRT simulations have significantly larger compute times, 857 mins versus 8.6 mins for $N_p = 1289$, compared to the correlations is because of the necessity to launch and track ~$10^6$ rays for intersections with each particle. Hence, view factor correlations can be about 100 times faster than MCRT simulations while achieving reasonable accuracy even for high solid volume fractions. Overall, detecting up to 5 nearest shading particles in particle-particle view factor computations using correlations developed in this study results in combined benefits of prediction accuracy and computational efficiency.

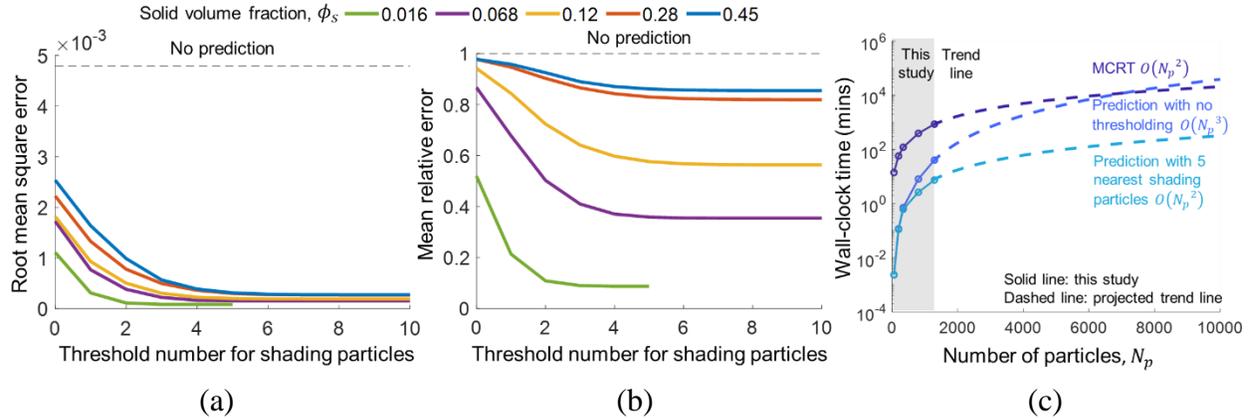

**Fig. 11.** Effects of thresholding the number of shading particle detection on prediction errors of particle-particle view factors quantified by (a) root mean square error (RMSE) and (b) mean relative error (MRE); (c) wall-clock time (mins) and algorithm complexity as a function of the number of particles for Monte Carlo ray tracing simulations and predictions with and without thresholding; actual data (solid) for compute times up to 1289 particles is projected in the trend lines (dashed) for up to $10^4$ particles.

## 4. Summary & Conclusion

In this study, we have developed data-driven correlations of particle-particle and particle-wall radiative view factors as a function of spatial locations and size of particles and wall surfaces. Compared to prior work, unique advancements in this study are the development of physically interpretable correlations, and explicitly checking and accounting for shading effects of neighboring particles, which becomes especially significant for large solid volume fractions.

Monte Carlo ray tracing simulations were performed on a random packed bed of monodisperse spheres with varying solid volume fractions of 0.016–0.45 to determine particle-particle and particle-wall view factors. This provides training and validation datasets to develop view factor correlations. Without any shading effects, the particle-particle view factor is governed by an inverse squared-relationship with the dimensionless distance, $d^*$, which is the ratio of the inter-particle separation distance to the particle diameter. The best-fit function to predict maximum particle-particle view factor (i.e., without any shading) involves 5 feature variables in the form of $d^{*m}$, where $m$ varies from -4 to 0. To account for shading effects, shading factors for individual particles are added based on their respective viewing angles, from which their product is deducted to account for overlaps in shadows cast by various particles. Predictions for shaded particle-particle view factors match ray tracing data with root mean square errors (RMSE) of 8.3e-5 and 2.7e-4, translating to 8.7% and 85% mean relative errors (MRE) for corresponding solid volume fractions of 0.016 and 0.45. Higher solid volume fraction results in larger errors due to the complexity of predicting shading effects by neighboring particles and the high density of view factors less than 1e-4, which particularly amplifies relative errors. For particle-wall view factor predictions, a correction factor, which is a quadratic function solely based on particle-wall normal distance, quantifies the effect of shading by intermediate particles. Predictions for shaded particle-wall view factors exhibit an excellent match with ray tracing data with RMSE and MRE values of 0.021 and 87% for the largest solid volume fraction modeled of 0.45.

View factor correlations developed in this study result in noteworthy accuracy, but accounting for all possible shading surfaces between a pair of particles will be computationally limiting, especially for large systems with many millions of particles. To this end, the effects of thresholding the number of shading particles was analyzed. Results show that accounting for shading effects by the nearest 5 neighboring particles can balance prediction accuracy with computational efficiency.

Overall, the radiative view factor correlations developed in this study exhibit the appeal of simplicity while being accurate and time efficient compared to ray tracing techniques to determine pairwise view factors with shading effect considered in particulate media. Additionally, correlations developed here enable a discrete approach to compute radiative fluxes on particle surfaces, and provide an opportunity to integrate radiative heat transfer in discrete/Lagrangian calculations of contact-driven forces and conductive heat transfer for flowing particles.

## Acknowledgments

Chen and Bala Chandran were partially supported by the Donors of the American Chemical Society Petroleum Research Fund (ACS-PRF, 62639-DNI9), and by the National Science Foundation under Grant No. 2144184, in this research. Chen was in-part also funded by the Rackham International Student Fellowship. Additionally, the authors acknowledge the financial support from the Department of Mechanical Engineering and the College of Engineering startup funds at the University of Michigan.